\documentstyle[prl,aps,psfig,multicol]{revtex}
\begin{document}
\draft

\title{Phases of the two-band model of spinless fermions in one dimension}
\author{Urs Ledermann and Karyn Le Hur}
\address{Theoretische Physik, Eidgen\"{o}ssische Technische Hochschule,
CH-8093 Z\"{u}rich, Switzerland}

\date{\today}
\maketitle

\begin{abstract}
{We study the two-band model of spinless fermions in one dimension for
weak repulsive interactions. In this case, the model is 
equivalent to the weakly interacting spinless two-leg ladder.
We obtain non-universal analytic expressions for the power-law decays
$\propto x^{-\gamma}$ of the charge-density- and $\propto x^{-1/\gamma}$ of
the superconducting pairing correlation functions. Leading order in the doping
away from half filling $\delta$ and $t_{\perp}/t$ we find
$\gamma=1+(\pi^{2}/8)(t_{\perp}/t)^{2}\delta^{2}$ ($t$ and $t_{\perp}$ are
the hopping terms along- and between the chains), such that superconducting
pairing correlations dominate. We furthermore show that the transition from the
superconducting phase to the usual one-dimensional (Luttinger) metal occurs
via a mixed phase, where superconducting pairs are formed in the bonding band only.
We give the phase diagram as a function of temperature and doping.}
\end{abstract}


\begin{multicols}{2}

\narrowtext

\section{Introduction}
One-dimensional (1D) electron systems have attracted much attention
over the past decade(s). While 3D systems are well described by the
Fermi liquid theory, for 1D systems this is not the case. The low-energy physics
is rather the one of a \emph{Luttinger liquid} (LL) \cite{bib:Luttinger,bib:Haldane},
where the excitations are collective zero sound modes. The Hubbard-
and the $t-J$ chains provide so far the most reasonable description of 1D
(super)conductors \cite{bib:Jerome,bib:Anderson}. 

The extension to quasi-one-dimensional systems is obtained by coupling~$N$
(Hubbard, $t-J$) chains to form a $N$-leg ladder. The interest in doped ladders started,
when the possibility of superconductivity in two-leg ladder materials was proposed
\cite{bib:DRS,bib:RGS,bib:DagRice}. For the Hubbard- and the $t-J$ two-leg ladders
different numerical methods have shown that $d$-wave like superconducting
correlations indeed dominate~\cite{bib:NWS,bib:Hal,bib:KKA}.
Experimentally, superconductivity has up to now only been observed
in a two-leg ladder material under high pressure \cite{bib:UNA}.

The Hubbard chain is exactly solvable for all values of the hopping parameter
$t$ and the on-site repulsion $U$ \cite{bib:LW}, and the $t-J$ chain at the
supersymmetric point, $t=J$~\cite{bib:BB}. Analytic works for ladders were made
for particular limits: For \emph{weakly coupled} chains (small interchain hopping
$t_{\perp}\ll U,t$) \cite{bib:Fabrizio,bib:KR,bib:Schu}, and for stronger
coupled chains, but \emph{small interactions} (small on-site repulsion
$U\ll t_{\perp},t$, such that the $N$-leg ladder is equivalent to a
$N$-band model)~\cite{bib:Fabrizio,bib:BF,bib:LBF,bib:LBFso8}.
Finally, \emph{at- and very close to half filling}, where recent works have shown
how to map the (weakly interacting) two-leg ladder on the Gross-Neveu model
\cite{bib:LBFso8} (for an application, see Ref. \cite{bib:KLLS}) and the
$t-J-J_{\perp}$ model on the $XXZ$ spin chain~\cite{bib:Schulzxxz}.

In order to determine whether the system exhibits superconductivity or not,
it is necessary to calculate the fluctuations (correlation functions) for
the charge-, and spin-density and for the superconducting (SC) pairing.
In the two-leg ladder, the decay of the charge density
correlation function is $\propto x^{-2K}$ and of the pairing correlation function
$\propto x^{-1/(2K)}$, where $K$ is the Luttinger liquid parameter (LLP) being
one in the noninteracting case, $K=1$. For sufficiently weak interactions
pairing correlations thus dominate which is interpreted as the appearance
of \emph{superconductivity}~\cite{bib:Schu,bib:BF}.

\emph{Spinless fermions} are usually studied as a model for the physical more
interesting (but more difficult) case of spin-$1/2$ fermions --- in particular
when focusing on the interchain hopping $t_{\perp}$ \cite{bib:Nerse} or the
metal-insulator transition \cite{bib:SLFInfDim}.
Physically,
spinless fermions can be considered as completely polarized spin-$1/2$ fermions in
a (high) magnetic field. For \emph{weak interactions}, a two-leg ladder of spinless
fermions is conveniently mapped on a two-band model. Without making the link
to ladders, such a model has
been treated in Ref. \cite{bib:MuEm}. More recent works have studied two
\emph{weakly coupled} chains, $t_{\perp}\ll t$ \cite{bib:Fabrizio,bib:Nerse,bib:Gogolin}. The authors
found that the LL fixed point of the two separated chains is \emph{unstable} upon weak
coupling. But in contrast to the spin-$1/2$ case, in a two-leg ladder of spinless fermions, the
decay of the charge density correlation function is $\propto x^{-K}$ and that of the SC pairing is
$\propto x^{-1/K}$, and superconductivity does not occur for
\emph{purely repulsive} interactions (in the $t-V-U$ model $V>0$ and $U>0$) and $t_{\perp}\ll t$
\cite{bib:Nerse,bib:Orignac}.

In this paper, we study the low-energy physics of the spinless two-leg
ladder for \emph{weak repulsive interactions and finite interchain hopping}
$t_{\perp}\sim t$. We obtain a non-universal analytic exponent for the
power-law decays $\propto x^{-\gamma}$ of the charge density- and
$\propto x^{-1/\gamma}$ of the SC pairing correlation functions.
The phases as function of the temperature $T$ and the doping away from
half filling $\delta$ (for spinless fermions, $0\leq\delta\leq 0.5$)
are shown in Fig.~1. At half filling the ladder is insulating; upon doping away
from half filling, we come to a phase with one gapless mode, which exhibits due to
the finite $t_{\perp}$ dominant SC correlations with \emph{inter}chain pairing.
When the doping is increased, this phase undergoes a transition to a phase with two
gapless modes, where first SC correlations coexist with charge density correlations.
The SC pairing correlations finally disappear completely at high doping, where
the ladder becomes a 1D Luttinger metal. We thus have a nontrivial crossover from
the SC state to the normal conducting one.

\begin{figure}[t]
 \centerline{
    \psfig{file=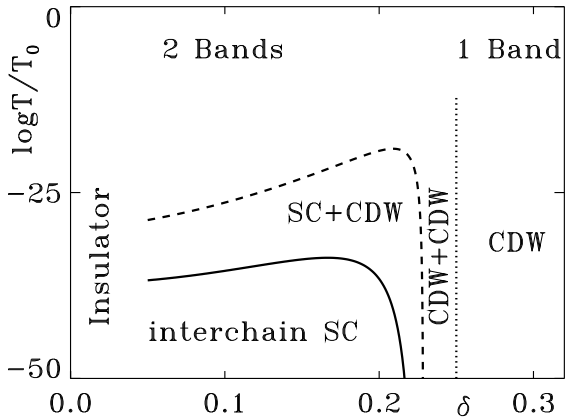,width=8cm,height=6cm}}
  \vspace{2mm}
  \caption{Phase diagram of the spinless two-leg ladder for $t_{\perp}/t=1$ and $U/t=0.2$.
   The doping away from half filling is $\delta$ and $T$ is the temperature (the $T_{0}$
   is a high temperature cutoff).
   The solid line shows the crossover to a superconducting (SC)
   phase with \emph{inter}chain pairing, and the dashed line displays where the
   crossover from a metallic phase with charge density wave (CDW) excitations
   to a mixed phase with coexistence of SC and CDW excitations occurs. The dotted
   line separates the region where both bands are partially filled, from the
   region where one band is empty.
   At half filling, the ladder is insulating. }
  \label{f:PhaseDiag}
\end{figure}

To determine the low-energy properties, we make use of the
renormalization group (RG) method \cite{bib:Shankar} and of bosonization
\cite{bib:Haldane,bib:SchulzRev}. 

In section~II, we introduce the Hamiltonian and prove the existence of a
gapless mode using a modified version of the Lieb-Schultz-Mattis (LSM) theorem
\cite{bib:LSM,bib:YOA,bib:GHR}. In section~III, we give the RG-equations (RGEs) and the
scaling of the couplings.
In section~IV, using bosonization the phases are
derived and discussed. Appendix~A contains the analytic solution of the one-loop
RGEs. In Appendix~B bosonization is explained and in Appendix~C
the correlation functions are calculated.

\section{The two-leg Ladder}
The noninteracting two-leg ladder of spinless fermions is given by the Hamiltonian,
\begin{eqnarray}
  H_{0}=-t\sum_{x,j} d_{j}^{\dagger}(x+1)d_{j}(x)
    +{\mathrm H.c.}
    \nonumber\\
    -t_{\perp}\sum_{x}d_{1}^{\dagger}(x)d_{2}(x)
    +{\mathrm H.c.},
\end{eqnarray}
where $t$ and $t_{\perp}$ are the hopping amplitudes along- and between the chains
and $d_{j}^{\dagger}(x)$ creates a fermion in the chain $j$ at the rung $x$.
We are going to consider \emph{small repulsive interactions} $0<U\ll t,t_{\perp}$.
In this limit, it is a good approach to diagonalize first $H_{0}$ by a
canonical transformation,
\begin{equation}
  \Psi_{j}(x)=\frac{1}{\sqrt{2}}\left(d_{1}(x)\pm d_{2}(x)\right).
\end{equation}
Going over to the momentum space, we find a decoupling into two bands
(we set the lattice parameter equal one),
\begin{equation}
  H_{0}=\sum_{j=1,2}\int dk\epsilon_{j}(k)\Psi_{j}^{\dagger}(k)\Psi_{j}(k),
  \label{eq:H02leg}
\end{equation}
where the dispersion relations are
\begin{equation}
  \epsilon_{j}(k)=\mp t_{\perp}-2t\cos(k).
  \label{eq:Disprel}
\end{equation}
Band 1 is the bonding- and band 2 the antibonding band. By analogy with the 2D case,
the associated transverse momenta are denoted as $k_{\perp}=0,\pi$. Since we are discussing
only the low-energy physics, we linearize $\epsilon_{j}$ around the Fermi
momenta $\pm k_{Fj}$ resulting in Fermi velocities $v_{j}=2t\sin(k_{Fj})$.
For the operator $\Psi_{j}$ at $\pm k_{Fj}$, we write $\Psi_{R/Lj}(k)=\Psi_{j}(\pm k_{Fj}+k)$.

In the small $U$ limit, it is a good approximation to take for $k_{Fj}$ the values
obtained in the noninteracting system \cite{bib:BF}. 
The definitions of the band filling $n=(k_{F1}+k_{F2})/(2\pi)$ and
the chemical potential, $\mu=\epsilon_{1}(k_{F1})=\epsilon_{2}(k_{F2})$
allow to calculate $k_{Fj}$ and $v_{j}$ as a function of $n$, $t$, and $t_{\perp}$,
\begin{equation}
  v_{1,2}=2t\sin\left[\pi n\pm\arcsin\left(\frac{t_{\perp}}{2t\sin(\pi n)}\right)\right].
\end{equation}
For spinless ladders, the band filling is $0\leq n\leq 1$ and the hole doping away
from half filling is $\delta=0.5-n$. The effect of the interchain hopping
$t_{\perp}$ is thus included in the velocities $v_{j}$ \cite{bib:BF}. We like to point out
that the velocities are \emph{not} equal for finite $t_{\perp}$,
\begin{equation}
  v_{1}-v_{2}=\frac{2t_{\perp}}{\tan(\pi n)}.
\end{equation}
This is \emph{different} to previous treatments of the spinless two-leg ladder
\cite{bib:Nerse}. We will see that this difference in the velocities
has the remarkable effect of driving the system to a SC state for repulsive
interactions, $U>0$. We note that both bands are partially filled,
when $t_{\perp}<2t\sin(\pi n)^{2}$.

We do not consider the half-filled case, $k_{F1}+k_{F2}=\pi$ ($v_{1}=v_{2}$), where
the ladder is insulating, allowing us to
neglect umklapp processes (we also exclude the particular points $k_{Fj}=\pi/2$).
Including all interactions allowed by symmetry 
(leaving away completely chiral one's), in momentum space, the Hamiltonian
is given by $H=H_{0}+H_{\rm Int}$, where
\begin{equation}
  H_{0}=\sum_{j=1,2}v_{j}\int dk k\left[\Psi_{Rj}^{\dagger}(k)\Psi_{Rj}(k)
    -\Psi_{Lj}^{\dagger}(k)\Psi_{Lj}(k)\right],
\end{equation}
and
\begin{eqnarray}
  H_{\rm Int}=\int dk_{1}dk_{2}dk_{3}dk_{4}\delta(k_{1}+k_{3}-k_{2}-k_{4})
  \nonumber\\
    \times\left[g_{1}\Psi_{R1}^{\dagger}(k_{1})\Psi_{R1}(k_{2})
    \Psi_{L1}^{\dagger}(k_{3})\Psi_{L1}(k_{4})
    +g_{2}(1\leftrightarrow 2)\right.
  \nonumber\\
  +g_{x}\left(\Psi_{R1}^{\dagger}(k_{1})\Psi_{R1}(k_{2})
    \Psi_{L2}^{\dagger}(k_{3})\Psi_{L2}(k_{4})
    +1\leftrightarrow 2\right)
  \nonumber\\
  \left.+g_{t}\left(\Psi_{R1}^{\dagger}(k_{1})\Psi_{R2}(k_{2})
    \Psi_{L1}^{\dagger}(k_{3})\Psi_{L2}(k_{4})+1\leftrightarrow 2\right)\right].
\end{eqnarray}
The bare values of the couplings are chosen as $g_{1}=g_{2}=g_{x}=g_{t}=U>0$.
We will see, that the $g_{t}$ interaction (pair hopping of left/right going quasiparticles from
band one to band two) is the most relevant in determining the low-energy physics.

Next, we apply the generalized LSM theorem (see Refs. \cite{bib:LSM,bib:YOA,bib:GHR}),
to the Hamiltonian $H$ in order to show the existence of gapless modes.
The particular symmetry of the Hamiltonian $H$ allows to define ``twist operators''
for left/right going fermions separately, i.e.,
\begin{equation}
  U_{R}=\exp\left(2\pi i\sum_{x,j}\frac{x}{L}\Psi_{Rj}^{\dagger}(x)\Psi_{Rj}(x)\right),
\end{equation}
where $L$ is the length of the ladder (and similarly $U_{L}$ for the left going fermions $\Psi_{Lj}$).
Commuting $U_{R/L}$ with the translation operator $T$ in real space, we obtain
$TU_{R}T^{-1}=e^{-4\pi i\nu_{R}}U_{R}$. The chiral symmetry $n=2\nu_{R}=2\nu_{L}$
proves then the existence of gapless (charge
density) excitations at a wavevector $2\pi n=k_{F1}+k_{F2}$, if $n$ is not an integer (remember
that we left away umklapp interactions in $H$ present at half filling; they break the individual
$U_{R/L}$ ``quasisymmetry'') \cite{bib:GHR}. We show, that this is in agreement with the results
obtained by bosonization (see section IV).
However, the LSM theorem does neither tell us the number of gapless modes nor whether finally
charge density- or superconducting excitations dominate.

\section{The RG-equations}
We give the RGEs resulting from the Hamiltonian $H$ and the flow of
the couplings depending on the ratio of the velocities $v_{1}/v_{2}$.
We find that for $v_{1}/v_{2}<7$, all couplings diverge, while
for $v_{1}/v_{2}>7$, $g_{t}\rightarrow 0$ and the rest remains of the order of $U$.

The model with the couplings $g_{1}$, $g_{2}$, and $g_{x}$ alone is exactly
solvable (by bosonization, see below) and in particular at a
\emph{RG fixed point}. Products of couplings
without at least one $g_{t}$ do therefore not appear in the RGEs
for $g_{1}$, $g_{2}$, and $g_{x}$. Including the one-loop exact results,
the particular form of the $g_{t}$ interaction then implies that
the RGEs (to all orders) have the following form,
\begin{eqnarray}
  \frac{dg_{1}}{dl}&=&-\frac{1}{2\pi v_{2}}g_{t}^{2}\left[1+{\mathcal O}(g_{\alpha}/t)\right]
  \nonumber\\
  \frac{dg_{2}}{dl}&=&-\frac{1}{2\pi v_{1}}g_{t}^{2}\left[1+{\mathcal O}(g_{\alpha}/t)\right]
  \nonumber\\
  \frac{dg_{x}}{dl}&=&\frac{1}{\pi(v_{1}+v_{2})}g_{t}^{2}\left[1+{\mathcal O}(g_{\alpha}/t)\right]
  \nonumber\\
  \frac{dg_{t}}{dl}&=&\frac{g_{t}}{\pi}\left[\frac{2g_{x}}{v_{1}+v_{2}}
    -\frac{g_{1}}{2 v_{1}}-\frac{g_{2}}{2 v_{2}} +{\mathcal O}(g_{\alpha}^{2}/t^{2})\right].
  \label{eq:RGEs2Band}
\end{eqnarray}
The energy (temperature) scale is related to $l$ by $T\sim te^{-l}$
and ${\mathcal O}(g_{\alpha}^{n}/t^{n})$ denotes higher order terms in the
couplings, $g_{\alpha}^{n}/t^{n}$ ($\alpha=1,2,x,t$).
The plus and minus signs result from particle-hole respectively
particle-particle diagrams. The one-loop RGEs have been derived
in Refs. \cite{bib:MuEm,bib:Fabrizio}, but the authors have not noted the particular
form to \emph{all} orders.

The exact solution of (\ref{eq:RGEs2Band}) to one-loop order is given in
Appendix~A. We obtain that for comparable ratios of the velocities,
$1/7<v_{1}/v_{2}<7$, all couplings diverge at the same scale,
while for other ratios, $g_{t}$ scales to zero and $g_{1}$, $g_{2}$ and $g_{x}$
stay of the order of $U$. The stability of the $g_{t}=0$ fixed point follows from the particular
form of the RGEs to all orders (\ref{eq:RGEs2Band}). Since all couplings
(to all orders) are multiplied at least once with $g_{t}$, this fixed point is
\emph{stable} for $g_{t}\rightarrow 0$ (any higher order term of the form $g_{1,2,x}^{3}$
would drive the system to another fixed point).

\section{Bosonization and Phases}
Using bosonization and the above RG results, we derive the low-energy phases (for
an overview, see Fig.~1). When doping the half filled ladder, a SC phase
with interchain pairing arises (up to $v_{1}/v_{2}<7$), see section~IV.A.
In section IV.B, we discuss the $g_{t}=0$ phase, present well away from half filling,
$v_{1}/v_{2}>7$, and exhibiting two gapless modes. In section IV.C, we finally
study the transitions and crossovers between the different phases.

Bosonizing the Hamiltonian $H$, we obtain (see Appendix B),
\begin{eqnarray}
  H=\int dx \sum_{j=1,2}\left[\frac{v_{j}}{2}+\frac{g_{j}}{4\pi}\right]
    (\partial_{x}\Phi_{j})^{2}
    +\left[\frac{v_{j}}{2}-\frac{g_{j}}{4\pi}\right]\Pi_{j}^{2}
  \nonumber\\
   +\frac{g_{x}}{2\pi}\left[\partial_{x}\Phi_{1}\partial_{x}\Phi_{2}-\Pi_{1}\Pi_{2}\right]
    -\frac{g_{t}}{(2\pi\alpha)^{2}}\cos\left[\sqrt{4\pi}(\theta_{1}-\theta_{2})\right].
  \label{eq:Bos12}
\end{eqnarray}
A flow to strong coupling of $g_{t}$ results (classically) in a ``pinning'' of
$\theta_{1}-\theta_{2}=0$ in order to minimize the energy, and a single gapless
mode, while for $g_{t}\rightarrow 0$, two gapless modes are present.

\subsection{The interchain-pairing SC phase}
We show that as a result of a finite interchain hopping $t_{\perp}$, doping the
half filled ladder, we obtain a phase, where SC correlations dominate. The pairing
takes place between left (right) going particles in chain one and right (left) going
particles in chain two. For $t_{\perp}\ll t$, we recover previous results \cite{bib:Nerse}.

When doping the half filled ladder (up to a ratio  $v_{1}/v_{2}<7$), the couplings grow
and eventually diverge. However, one should not overinterpret this divergence.
For small $U$, the combination $\theta_{1}-\theta_{2}$ in (\ref{eq:Bos12}) can fluctuate
at high temperatures, but for growing $g_{t}$ (i.e., decreasing temperature),
these fluctuations become suppressed
at the scale where $g_{t}(T_{c})\sim t$. The $\theta_{1}-\theta_{2}=0$ then
leads to a coherence between the two bands and is interpreted as a crossover to a new phase.
Since $T_{c}\sim te^{-l_{c}}\ll t$ is a very low temperature ($l_{c}\propto t/U$), the change of the
couplings as a function of temperature for $T<T_{c}$ is weak as long as we keep away from
the point of divergence, i.e., within the validity of the one-loop RGEs.
It is then a common practice
to set $g_{t}\sim t$ as a strong coupling value (and similarly, for the other couplings).

Using the canonical transformation
$\Phi_{\pm}=(\Phi_{1}\pm\Phi_{2})/\sqrt{2}$ and $\Pi_{\pm}=(\Pi_{1}\pm\Pi_{2})/\sqrt{2}$,
the Hamiltonian (\ref{eq:Bos12}) takes the form
\begin{equation}
  H=H_{B}+H_{\mathrm SG}+H_{\mathrm mix},
  \label{eq:Hpm}
\end{equation}
where $H_{B}$ is the Hamiltonian of a massless boson,
\begin{equation}
  H_{B}=\int dx\frac{u_{+}}{2}\left[\frac{1}{K_{+}}
    (\partial_{x}\Phi_{+})^{2}+K_{+}\Pi_{+}^{2}\right],
\end{equation}
and $H_{\mathrm SG}$ is the sine-Gordon Hamiltonian,
\begin{eqnarray}
  H_{\mathrm SG}=\int dx\left\{\frac{u_{-}}{2}\left[\frac{1}{K_{-}}
    (\partial_{x}\Phi_{-})^{2}+K_{-}\Pi_{-}^{2}\right]\right.
  \nonumber\\
    \left.-\frac{g_{t}}{(2\pi\alpha)^{2}}
    \cos\left[\sqrt{8\pi}\theta_{-}\right]\right\},
  \label{eq:HSG}
\end{eqnarray}
and finally, $H_{\mathrm mix}$ is a mixing term,
\begin{equation}
    H_{\mathrm mix}=\int dx\left(v_{-}^{c}\partial_{x}\Phi_{+}\partial_{x}\Phi_{-}
    +v_{-}^{p}\Pi_{+}\Pi_{-}\right).
  \label{eq:Hmix}
\end{equation}
The velocities $u_{\pm}$, $v_{-}^{c,p}$, and the LLP $K_{\pm}$ are
given by
\begin{equation}
  u_{\pm}=\sqrt{\left(\frac{v_{1}+v_{2}}{2}\right)^{2}-
    \left(\frac{g_{1}+g_{2}\pm 2g_{x}}{4\pi}\right)^{2}},
\end{equation}
\begin{equation}
  v_{-}^{c,p}=\frac{v_{1}-v_{2}}{2}\pm\frac{g_{1}-g_{2}}{4\pi},
\end{equation}
and
\begin{equation}
  K_{\pm}=\sqrt{\frac{2\pi(v_{1}+v_{2})-(g_{1}+g_{2}\pm 2g_{x})}
    {2\pi(v_{1}+v_{2})+(g_{1}+g_{2}\pm 2g_{x})}}.
  \label{eq:Krho}
\end{equation}
The mixing term hinders for $v_{1}-v_{2}\neq 0$ a (simple) analytic
solution of the (classical) equations of motion. Since the $\theta_{-}$ field is pinned, the current
density takes the form $j=u_{+}K_{+}\Pi_{+}$. At half filling the $\Phi_{+}$ field is also pinned
resulting in $j=0$ and an insulating phase \cite{bib:MuEm}.

Next, we discuss the correlation functions. For repulsive interactions, the charge density- and
SC pairing fluctuations with the most divergent susceptibilities are the following ones:
The CDW (at a wavevector $k_{F1}+k_{F2}$, for a comparison, see section II) is given by the operator
\begin{eqnarray}
  O_{\mathrm CDW}&=&d_{R1}^{\dagger}d_{L1}-d_{R2}^{\dagger}d_{L2}=
    \Psi_{R2}^{\dagger}\Psi_{L1}+\Psi_{R1}^{\dagger}\Psi_{L2}
  \nonumber\\
  &&\propto
    \exp\left(i\sqrt{2\pi}\Phi_{+}\right)\cos\left(\sqrt{2\pi}\theta_{-}\right),
  \label{eq:CDWO}
\end{eqnarray}
and the SC fluctuations by
\begin{eqnarray}
  O_{\mathrm SC}&=&d_{R1}d_{L2}+d_{R2}d_{L1}=\Psi_{R1}\Psi_{L1}-\Psi_{R2}\Psi_{L2}
  \nonumber\\
  &&\propto
    \exp\left(i\sqrt{2\pi}\theta_{+}\right)\cos\left(\sqrt{2\pi}\theta_{-}\right),
  \label{eq:PFO}
\end{eqnarray}
where the $d_{R/Lj}$ are the annihilation operators for the fermions in \emph{chain} $j$ \cite{bib:Error}.
The operator $O_{\mathrm CDW}$ represents an antisymmetric charge density wave and $O_{\mathrm SC}$
superconductivity with \emph{inter}chain pairing --- previously called
$d$-wave like due to the antisymmetry with respect to the bonding- and antibonding band
\cite{bib:Orignac}. However, the operator $O_{\mathrm SC}$ has \emph{odd} parity,
$O_{\mathrm SC}(-x)=-O_{\mathrm SC}(x)$, which one associates rather with $p$-wave like
superconductivity in each band (see Appendix C.1).

We obtain for the charge density correlation function (for a derivation, see Appendix~C),
\begin{equation}
  \left\langle O_{\mathrm CDW}^{\dagger}(x)O_{\mathrm CDW}(0)\right\rangle\propto x^{-\gamma}
\end{equation}
and for the SC pairing correlation function
\begin{equation}
  \left\langle O_{\mathrm SC}^{\dagger}(x)O_{\mathrm SC}(0)\right\rangle\propto x^{-1/\gamma},
\end{equation}
where the exponent is 
\begin{equation}
  \gamma=\frac{K_{+}}{1-\frac{K_{+}K_{-}}{2 u_{+}u_{-}}\left(v_{-}^{c}\right)^{2}}.
  \label{eq:GamExp}
\end{equation}
\emph{As a result, the finite} $t_{\perp}\propto v_{-}^{c}$
\emph{leads to} $\gamma>1$, \emph{implying that superconducting pairing correlations
dominate}.

The increase of $\gamma$ can be understood as follows. The pinning of $\theta_{-}$
allows us to set $\Pi_{-}=0$. The only coupling is then between
the $\Phi_{\pm}$ fields. The $\Phi_{-}$ field fluctuates strongly and
affects the $\Phi_{+}$ correlation function with additional
fluctuations thus increasing $\gamma$ and stabilizing the superconductivity.

We furthermore find that in the low-energy regime,
$T\rightarrow 0$ ($l\rightarrow\infty$), the LLP $K_{+}$ becomes bigger than one.
In detail, whether $K_{+}$ is bigger or smaller than one depends on the following
sum (see (\ref{eq:Krho}))
\begin{equation}
  g_{1}+g_{2}+2g_{x}=\left( 3+\frac{v_{1}^{2}+v_{2}^{2}}{2v_{1}v_{2}}\right)U
    -\frac{(v_{1}-v_{2})^{2}}{2v_{1}v_{2}}g_{x}.
\end{equation}
Since $g_{x}$ increases for decreasing temperature, the above sum becomes
negative and $K_{+}>1$.

\begin{figure}[t]
 \centerline{
    \psfig{file=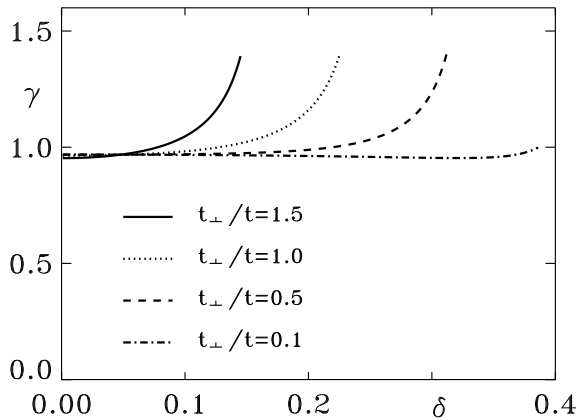,width=8cm,height=6cm}}
  \vspace{2mm}
  \caption{The exponent $\gamma$ of the charge density correlation function,
    $\propto x^{-\gamma}$, is shown for $U/t=0.2$ and different ratios of
    $t_{\perp}/t$. While for small ratios $t_{\perp}\ll t$, CDW dominate for almost
    all physical values of the doping $\delta$, comparable ratios $t_{\perp}\sim t$
    favor superconductivity already at lower doping.
    The SC correlations are strongest, when approaching the transition to the mixed
    SC+CDW phase, where the chemical potential is close to the bottom of the antibonding band,
    ($\gamma\rightarrow 1.5$ for $v_{1}/v_{2}\rightarrow 7$).}
\end{figure}

It is instructive to expand (\ref{eq:GamExp}) leading order in $g_{x}/t$ and $t_{\perp}/t$
(we rewrite $g_{1}$ and $g_{2}$ in terms of $g_{x}$ and neglect $U\ll t$), 
\begin{equation}
  \gamma=1+\frac{1}{2}\left(\frac{v_{1}-v_{2}}{v_{1}+v_{2}}\right)^{2}
    \left(1+\frac{v_{1}+v_{2}}{2\pi v_{1}v_{2}}g_{x}\right).
  \label{eq:GamExpSimpl}
\end{equation}
Neglecting the $g_{x}\sim t$ due to the relatively small prefactor, we rewrite
(\ref{eq:GamExpSimpl}) in terms of physical quantities,
\begin{equation}
  \gamma=1+\frac{1}{8}\left(\frac{t_{\perp}}{t}\right)^{2}\frac{\cot(\pi n)^{2}}
    {\sin(\pi n)^{2}-(t_{\perp}/2t)^{2}}.
\end{equation}
Leading order in the doping $\delta=0.5-n$, we then find the particular simple form,
\begin{equation}
  \gamma=1+\frac{\pi^{2}}{8}\left(\frac{t_{\perp}}{t}\right)^{2}\delta^{2}.
  \label{eq:GammaApprox}
\end{equation}
The $\delta^{2}$ and the $t_{\perp}^{2}$ reflect the 
$v_{1}\leftrightarrow v_{2}$ symmetry. Comparing (\ref{eq:GammaApprox}) with the
usual exponent $K$ of a LL, we deduce that the $t_{\perp}$ term produces an effective
attraction between particles of chain one and chain two.

The exponent $\gamma$ as a function of the doping $\delta$ is shown in Fig.~2 for different
ratios $t_{\perp}/t=0.1,\ldots,1.5$ (we have set $g_{x}=t$) \cite{bib:GExpl}.
It becomes smaller than one when approaching half filling,
$\gamma\approx K_{+}\leq 1$ (a precursor of the insulating state), and 1.5 near
the transition to the mixed phase.
The insulating state at half filling thus evolves (over a phase with weak CDW correlations; this
depends on the choice of $g_{x}\sim t$) to a SC
phase with strong correlations, where the chemical potential lies near the bottom of the antibonding
band ($v_{1}/v_{2}\approx 7$) \cite{bib:NBSZ}.
A comparison of Fig.~2 with Fig.~1 shows that the SC correlations are largest where
the crossover temperature is highest.

For small $t_{\perp}/t$ we recover previous results: the bosonized Hamiltonian goes
over to the one studied in Refs. \cite{bib:Nerse,bib:Orignac} and the exponent $\gamma$ stays
smaller than one, $\gamma\approx K_{+}\leq 1$, and charge density correlations dominate (for almost
all physical values of the doping $\delta$).

\subsection{The mixed SC+CDW phase}
For $v_{1}/v_{2}>7$, the fixed points of the couplings are such that $g_{t}=0$,
and $g_{1}$, $g_{2}$, and $g_{x}$ are of the
order of $U$ (see Appendix~A). We diagonalize the Hamiltonian (\ref{eq:Bos12}) for
$g_{t}=0$ using a current representation of the fields (for a comparison, see Appendix~B).
Defining ${\mathbf J}=(J_{R1},J_{L1},J_{R2},J_{L2})$ and
\begin{equation}
  M=
  \left(\begin{array}{cccc}
        v_{1}         & g_{1}/2\pi  & 0              & g_{x}/2\pi  \\
        g_{1}/2\pi  & v_{1}         & g_{x}/2\pi & 0               \\
        0               & g_{x}/2\pi  & v_{2}        & g_{2}/2\pi  \\
        g_{x}/2\pi  & 0               & g_{2}/2\pi & v_{2}
  \end{array}\right),
\end{equation}
the Hamiltonian (\ref{eq:Bos12}) can be written as
$H=\pi\int dx {\mathbf J}^{T}M{\mathbf J}$ and
the new LLP and velocities are determined by the eigenvalues of M,
which are
\begin{eqnarray}
  \frac{v_{1}+v_{2}}{2}+\frac{g_{1}+g_{2}}{4\pi}\pm\frac{D_{+}}{2},
  \nonumber\\
  \frac{v_{1}+v_{2}}{2}-\frac{g_{1}+g_{2}}{4\pi}\pm\frac{D_{-}}{2},
\end{eqnarray}
where
\begin{equation}
  D_{\pm}=\sqrt{\left(v_{1}-v_{2}\pm\frac{g_{1}-g_{2}}{2\pi}\right)^{2}
    +4\left(\frac{g_{x}}{2\pi}\right)^{2}}.
\end{equation}
Since $v_{1}-v_{2}\gg g_{x}$, the coupling $g_{x}$ does not contribute to $D_{\pm}$ leading
order in $U$. The velocities and the LLP have thus the same form as in two decoupled LL,
\begin{equation}
  u_{j}=\sqrt{v_{j}^{2}-\frac{g_{j}^{2}}{(2\pi)^{2}}}\;\;{\mathrm and}\;\;
    K_{j}=\sqrt{\frac{2\pi v_{j}-g_{j}}{2\pi v_{j}+g_{j}}}\approx 1-\frac{g_{j}}{2\pi v_{j}},
\end{equation}
where in our case the $g_{j}$ are scale dependent. The new basis is ``almost'' the old one,
$\tilde{\Pi}_{1}\approx\Pi_{1}+\epsilon\Pi_{2}$, where $\epsilon\sim U/t$ (and similarly for $\tilde{\Pi}_{2}$
and the fields $\tilde{\Phi}_{j}$) and the current density takes the usual form,
\begin{equation}
  j=u_{1}K_{1}\tilde{\Pi}_{1}+u_{2}K_{2}\tilde{\Pi}_{2},
\end{equation}
where the $\tilde{\Pi}_{j}$ obey the equations of motion
$\partial_{t}^{2}\tilde{\Pi}_{j}=u_{j}^{2}\partial_{x}^{2}\tilde{\Pi}_{j}$. It should be noted
that here the Drude coefficient $u_{j}K_{j}$ is not equal to that of the free
Fermi gas \cite{bib:SchulzRev}.

In a short range of ratios, $7<v_{1}/v_{2}<8$, the low temperature fixed point
of the couplings is such that $g_{1}<0$ and $g_{2}>0$ (for $v_{1}/v_{2}=7$, we
find $g_{1}=-0.57 U$ and $g_{2}=0.78 U$), implying $K_{1}>1$ and
$K_{2}<1$ --- \emph{a coexistence of pairing correlations in the bonding- with charge
density correlations in the antibonding band}. Rewriting the SC pairing operator
$O_{\mathrm SC1}=\Psi_{R1}\Psi_{L1}$ in terms of the chain operators $d_{R/Lj}$ shows,
that there are as well \emph{intra}- and \emph{inter}-chain SC pairs.
We interpret this coexistence as a precursor of the SC phase at $v_{1}/v_{2}<7$,
i.e., pairing correlations first become established in the bonding band
(at a momentum $k_{\perp}=0$), and then also in the antibonding
(at a momentum $k_{\perp}=\pi$) and phase coherence between the bands sets in for increasing $g_{t}$.
In Fig.~1, the dashed line shows, where $K_{1}>1$.

For ratios $v_{1}/v_{2}>8$, we find $g_{j}>0$ and in both bands
charge density fluctuations dominate, $K_{j}<1$
(for $v_{1}/v_{2}\gg 8$, the fixed points are $g_{1}\approx g_{2}\approx U$).
This corresponds to the usual 1D metallic (LL) phase. 

\subsection{Phase Transitions and Crossovers}
Next, we discuss the transitions (at $T=0$) and the crossovers between the different phases
as a function of doping and temperature (see Fig.~1). The transitions are accompanied
by non-analyticities in the compressibility $\kappa$ respectively in the coefficient $\lambda$
of the specific heat, $C=\lambda T$.

We interpret the energy-scale $T_{c}\sim te^{-l_{c}}$ ($l_{c}\propto t/U$)
at which the couplings become of the order of $t$ as the temperature, where the crossover
to the interchain SC phase takes place, see Fig.~1 (the high temperature cutoff $T_{0}$
is fixed but arbitrary). The crossover temperature $T_{c}$ increases upon doping, reaching a
maximum at a finite, $t_{\perp}/t$ dependent doping, and then decreases to zero when
approaching the transition point to the mixed SC+CDW phase, $v_{1}/v_{2}\approx 7$.

The compressibility $\kappa$ and the coefficient $\lambda$ of the specific heat in the mixed
phase have the same form as in two decoupled LL~\cite{bib:SchulzRev}, e.g.,
\begin{equation}
  \frac{\lambda}{\lambda_{0}}=\frac{1}{2}\left(\frac{v_{1}}{u_{1}}+\frac{v_{2}}{u_{2}}\right)
    \approx 1+\frac{1}{4}\left(\frac{g_{1}}{2\pi v_{1}}\right)^{2}
    +\frac{1}{4}\left(\frac{g_{2}}{2\pi v_{2}}\right)^{2},
\end{equation}
where $\lambda_{0}$ is the specific heat coefficient of the noninteracting system.
In the interchain SC phase, the main dependencies of $\kappa$ and $\lambda$ come from
$K_{+}$ and $u_{+}$, but similarly as for the correlation functions, there are corrections
in $v_{-}^{c}$. In any case, the values of $\lambda$ and $\kappa$ ``jump'' at $v_{1}/v_{2}=7$.

The SC+CDW phase then evolves into a CDW+CDW phase, where the crossover takes place at
$v_{1}/v_{2}\approx 8$ (at $T=0$). The band edge of the antibonding band
($v_{1}/v_{2}\rightarrow\infty$) cannot be treated by the present method, since the
dispersion relation is there quadratic, $\propto k^{2}$. However, choosing a sufficiently
small $U$, we can approach this point arbitrarily close. We thus conjecture that the CDW+CDW
phase can be extended up to this band edge.

When only one band is occupied, i.e., $v_{2}=0$, the
phase consists of a single LL (with CDW excitations); at the band edge of the
antibonding band, we then have a transition from a one-band CDW to a two-band CDW+CDW phase,
where for $U/t\rightarrow 0$, the coefficient $\lambda$ and the compressibility $\kappa$ diverge 
as $\lambda,\kappa\sim 1/k_{F2}$.

\section{Conclusions}
We have shown that a finite interchain hopping $t_{\perp}$ has the effect of driving 
the weakly interacting spinless two-leg ladder to a SC phase, where left (right) going
particles in chain one are paired with right (left) going particles in chain two. The
SC correlations are largest when the chemical potential is close to the bottom of the
antibonding band.
Between the SC phase at lower- and the CDW phase at higher doping, we have found a (new)
phase with coexistence of SC and CDW excitations. Two phase transitions take place
(for $T\rightarrow 0$): one from the interchain SC phase to the mixed SC+CDW phase and one
from the CDW+CDW to the (one band) CDW phase.

On the one hand, SC correlations are enhanced by the velocity difference
$v_{1}-v_{2}=2t_{\perp}/\tan(\pi n)$, but one the other hand, this difference suppresses
the coherence between the two bands, when the chemical potential is too close to the bottom
of the antibonding band. We argue that this sort
of competition also takes place in the spin-$1/2$ case \cite{bib:BF,bib:NBSZ} and generally for
(possible) superconductivity in $N$-band models.

\acknowledgments{We thank T.M. Rice for fruitful discussions and comments
throughout this work.}

\appendix

\section{Renormalization Group}
The renormalization group (RG) method is a controlled way of subsequently
eliminating (integrating out) high-energy modes in a given
Hamiltonian. While the noninteracting part $H_{0}$ is (usually) at a
\emph{RG fixed point}, couplings of the interacting part $H_{\mathrm{Int}}$
may grow or decrease under a RG transformation (i.e., when lowering the energy).

A subsequent (perturbative) elimination of 
high-energy modes in $H_{\mathrm{Int}}$ then results in RG equations, which
give the change of the coupling when lowering the energy, see Ref.~\cite{bib:Shankar}.

Next, we solve analytically the one-loop RGEs given in equation
(\ref{eq:RGEs2Band}). Keeping only the terms quadratic in the coupling constants,
we transform the set of the four differential equations (\ref{eq:RGEs2Band}) into one
differential equation for $g_{x}$,
\begin{equation}
  \frac{1}{B}\frac{dg_{x}}{dl}=\left(g_{x}-CU\right)^{2}+DU^{2},
  \label{eq:DGLgx}
\end{equation}
where
\begin{equation}
  B=\frac{4v_{1}v_{2}+(v_{1}+v_{2})^{2}}{2\pi v_{1}v_{2}(v_{1}+v_{2})},
\end{equation}
\begin{equation}
  C=\frac{2(v_{1}+v_{2})^{2}}{4v_{1}v_{2}+(v_{1}+v_{2})^{2}},
\end{equation}
and
\begin{equation}
  D=\frac{-v_{1}^{4}+6v_{1}^{3}v_{2}+6v_{1}^{2}v_{2}^{2}+6v_{1}v_{2}^{3}-v_{2}^{4}}
    {\left[4v_{1}v_{2}+(v_{1}+v_{2})^{2}\right]^{2}}.
\end{equation}
The solution of equation (\ref{eq:DGLgx}) is qualitatively different for $D<0$ and $D>0$.
For $D>0$, all couplings diverge, while for $D<0$, $g_{t}$ scales to zero and the
others remain of the order of $U$.

Solving the equation 
\begin{equation}
   x^{4}-6x^{3}-6x^{2}-6x+1=0,
\end{equation}
for $x=v_{1}/v_{2}$, we obtain the exact transition ratio $v_{1}/v_{2}$. For comparable
velocities,
\begin{equation}
  1/7\approx 0.14327\ldots<v_{1}/v_{2}<6.9798\ldots\approx 7,
\end{equation}
we find $D>0$ resulting in
\begin{equation}
  g_{x}(l)=U\left\{C+\sqrt{D}\tan\left[B\sqrt{D}Ul
    -\arctan\left(\frac{C-1}{\sqrt{D}}\right)\right]\right\}
\end{equation}
and
\begin{eqnarray}
  &g_{t}(l)=U\sqrt{\pi(v_{1}+v_{2})BD}
  \nonumber\\
  &\times\left\{1+\tan\left[B\sqrt{D}Ul
    -\arctan\left(\frac{C-1}{\sqrt{D}}\right)\right]^{2}\right\}^{1/2}.
\end{eqnarray}

For ratios $v_{1}/v_{2}>7(<1/7)$, we find $g_{t}\rightarrow 0$  and
\begin{equation}
  g_{x}\rightarrow(C-\sqrt{-D})U,
\end{equation}
for $l\rightarrow\infty$. Similar as $g_{x}$, the coupling constants $g_{1}$ and $g_{2}$
stay of the order of $U$ (e.g., $U<g_{x}<1.4U$). Since $g_{t}$ flows to zero and the
form of the RGEs to {\it all orders} (\ref{eq:RGEs2Band}) is such that couplings whatever the
order is, are always multiplied at least once with $g_{t}$, the fixed point is stable,
$dg_{\alpha}/dl\rightarrow 0$ ($\alpha=1,2,x,t$) for $l\rightarrow\infty$. For finite $U<v_{j}$, higher
order terms lead to a change of the transition ratio $v_{1}/v_{2}$.

The one-loop RGEs (\ref{eq:RGEs2Band}) have been used in Ref.~\cite{bib:Fabrizio} to treat
spin-$1/2$ fermions and the author has noted the two different flow regimes, but he has not
given the solutions of the RGEs.

\section{Bosonization}
Bosonization is a method of rewriting Dirac-fermion operators
$\Psi$ in terms of bosonic fields $\Phi$ and $\Pi$ satisfying the
commutation relation
\begin{equation}
  [\Phi(x),\Pi(y)]=i\delta(x-y).
\end{equation}
For fermions on a chain or ladder, bosonization applies in the \emph{continuum limit}.
A ``complete'' treatment of bosonization is given in Refs. \cite{bib:Haldane,bib:SchulzRev}.

For convenience, we introduce the dual field of $\Phi$,
\begin{equation}
  \theta(x)=\int_{-\infty}^{x}dx^{\prime}\Pi(x^{\prime}).
\end{equation}
For spinless fermions, the bosonization scheme is then as follows. We first decompose the fermionic
operator $\Psi$ into right- and left-movers $\Psi_{R}$ and $\Psi_{L}$,
\begin{equation}
  \Psi(x)=e^{ik_{F}x}\Psi_{R}(x)+e^{-ik_{F}x}\Psi_{L}(x).
\end{equation}
In terms of bosonic operators, the $\Psi_{R/L}$ take the form
\begin{equation}
  \Psi_{R/L}(x)=\frac{\eta_{R/L}}{\sqrt{2\pi\alpha}}
    \exp\left[i\sqrt{\pi}(\mp\Phi(x)+\theta(x))\right],
\end{equation}
where $\alpha$ is a cutoff parameter of the order of the
lattice constant and the $\eta_{R/L}$ are Majorana (``real'') fermionic operators
(usually called ``Klein factors''), which are necessary to fulfill the anticommutation
relations of the $\Psi_{R/L}$ fields.
Currents $J_{R/L}=\Psi_{R/L}^{\dagger}\Psi_{R/L}$ then become
\begin{equation}
  J_{L}+J_{R}=\frac{1}{\sqrt{\pi}}\partial_{x}\Phi\;\;{\mathrm and}\;\;
  J_{L}-J_{R}=\frac{1}{\sqrt{\pi}}\Pi.
\end{equation}
The generalization to $N$ species of spinless fermions is straightforward.

Fourier transforming the Hamiltonian $H$ of the spinless two-leg ladder
(two-band model) back to the $x$ space and using the above rules, we simply obtain the
bosonized Hamiltonian (\ref{eq:Bos12}). 

\section{Correlation functions}
For the spinless two-leg ladder in the bosonized representation (\ref{eq:Hpm}),
we calculate (equal time) charge density fluctuations
\begin{eqnarray}
  &\left\langle\left(e^{i\beta\Phi_{+}(x)}-\left\langle e^{i\beta\Phi_{+}(x)}\right\rangle\right)
    \left(e^{-i\beta\Phi_{+}(0)}-\left\langle e^{-i\beta\Phi_{+}(0)}\right\rangle\right)\right\rangle
  \nonumber\\
  &=\left\langle e^{i\beta\Phi_{+}(x)}e^{-i\beta\Phi_{+}(0)}\right\rangle
      -\left|\langle e^{i\beta\Phi_{+}(x)}\rangle\right|^{2}
\end{eqnarray}
and similarly, fluctuations of the SC pairing operator parameterized by the $\theta_{+}$ field
(for simplicity, we drop the pinned $\theta_{-}$ field). We make use of the path integral formalism,
where expectations values of operators are calculated by integration over all field configurations
weighted with the exponential of the classical action.

We first revisit the correlation functions of the free massless boson
and of the SG-model (see e.g. \cite{bib:Gogolin}).

\subsection{Massless boson and SG-model}
The nature of the phases strongly depends on the long-range behavior of
charge- and SC pairing correlation functions. In a single chain, such
correlation functions are determined by a single (interaction dependent)
parameter, usually denoted by~$K$ (Luttinger liquid parameter, LLP).

The bosonized action of a single chain of spinless fermions is in the
continuum limit the one of a
massless boson, i.e., it is gaussian in the field $\Phi$ allowing
for an analytic calculation of correlation functions. It is
convenient to perform a Wick rotation to imaginary time, $\tau=it$.
The action then reads,
\begin{equation}
  S=\frac{1}{2K}\int dxd\tau\Phi\left(\frac{1}{v}\partial_{\tau}^{2}
    +v\partial_{x}^{2}\right)\Phi,
  \label{eq:AFreeB}
\end{equation}
where $K$ is the interaction dependent LLP and $v$ the Fermi velocity.
The Green's function $G$ satisfying,
\begin{equation}
  -\left(\frac{1}{v}\partial_{\tau}^{2}+v\partial_{x}^{2}\right)
    G(x,\tau)=\delta(x)\delta(\tau),
\end{equation}
is given by
\begin{equation}
  G(x,\tau)=\frac{1}{4\pi}
    \ln\left(\frac{R^{2}}{x^{2}+v^{2}\tau^{2}+\alpha^{2}}\right),
  \label{eq:GFMB}
\end{equation}
where $\alpha$ is a short distance cutoff and $R$ the radius of the integration boundary
in the complex plane (we finally take $R\rightarrow\infty$, corresponding to
the usual thermodynamic limit). The one-point correlation function
is then equal zero $\left\langle e^{i\beta\Phi(x)}\right\rangle=0$.
Defining $O_{\mathrm CDW}=\Psi_{R}^{\dagger}\Psi_{L}$,
the long-range behavior of the (equal time) charge density
correlation function at $2k_{F}$ is 
\begin{equation}
  \left\langle O_{\mathrm CDW}^{\dagger}(x)O_{\mathrm CDW}(0)\right\rangle\propto x^{-2K}.
\end{equation}
Here, $\beta^{2}=4\pi$ (see above). The SC pairing operator is $O_{\mathrm SC}=\Psi_{R}\Psi_{L}$
and its correlation function
\begin{equation}
  \left\langle O_{\mathrm SC}^{\dagger}(x)O_{\mathrm SC}(0)\right\rangle\propto x^{-2/K}.
\end{equation}
A $K>1$ therefore implies dominant pairing-, while a $K<1$ results in dominant
charge density correlations. Note that the SC pairing operator has \emph{odd}
parity, $O_{\mathrm SC}(x)=-O_{\mathrm SC}(-x)$ (for spin-$1/2$ fermions, the parity is even).
For chiral fermions, the parity transformation is $\Psi_{L}(x)\rightarrow\pm\Psi_{R}(x)$ for
periodic (antiperiodic) boundary conditions.

The action of the SG-model has the form
\begin{equation}
  S=\frac{1}{2}\int dxd\tau\Phi\left(\partial_{\tau}^{2}+\partial_{x}^{2}\right)\Phi
    +g\cos(\beta_{0}\Phi),
\end{equation}
where we take $\beta_{0}^{2}<8\pi$. The ``pinning'' term $g\cos(\beta_{0}\Phi)$ renders
the one-point correlation function of the $\Phi$ field
a constant, $\left\langle e^{i\beta\Phi(x)}\right\rangle$=const$\neq 0$.
The two-point correlation function becomes a constant
at large distances (cluster decomposition principle),
\begin{equation}
   \left\langle e^{i\beta\Phi(x)}e^{-i\beta\Phi(0)}\right\rangle
   \rightarrow\left\langle e^{i\beta\Phi(x)}\right\rangle
   \left\langle e^{-i\beta\Phi(0)}\right\rangle={\mathrm const},
\end{equation}
for $x\rightarrow\infty$. The one-point correlation functions of the (unpinned) dual
field $\theta$ is equal zero and the two-point correlation function decays
exponentially.

\subsection{The spinless two-leg ladder}
Next, we calculate the correlation functions of
the spinless two-leg ladder making use of the above results. In our case, the
$\theta_{-}$ field is pinned. 

The action of the spinless two-leg ladder (resulting from the Hamiltonian (\ref{eq:Hpm}))
is invariant under the shift $\Phi_{+}(x)\rightarrow\Phi_{+}(x)+c$, where $c$ is any constant.
Similarly as it is the case for the action of the free massless boson (\ref{eq:AFreeB}),
we then obtain (for a comparison, see Ref. \cite{bib:Gogolin})
\begin{equation}
  \left\langle e^{i\sum_{j}\beta_{j}\Phi_{+}(x_{j})}\right\rangle=0,
\end{equation}
if $\sum_{j}\beta_{j}\neq 0$. In particular, the one-point correlation function
of the $\Phi_{+}$ field is equal zero, $\left\langle e^{i\beta\Phi_{+}(x)}\right\rangle=0$,
i.e., the $\Phi_{+}$ field is indeed a free (unpinned) field. The same holds
for the dual field $\theta_{+}$.
We conjecture that the mixing term (\ref{eq:Hmix}) for $v_{1}\neq v_{2}$ is an
\emph{analytic perturbation}.

Using the Green's function (\ref{eq:GFMB}), we carry out the integration over the
fields $\Pi_{+}$ and $\Phi_{+}$ and obtain for the equal time charge density correlation
function
\begin{equation}
  \left\langle e^{i\beta\Phi_{+}(x)}e^{-i\beta\Phi_{+}(0)}\right\rangle
    \propto x^{-2K_{+}\beta^{2}/4\pi}
    \left\langle e^{S_{\mathrm mix}^{\Phi}(x)}\right\rangle_{\mathrm SG}.
\end{equation}
The expression $S_{\mathrm mix}^{\Phi}$ depends (nonlocally) on the fields $\Phi_{-}$, $\Pi_{-}$
vanishing for $v_{1}=v_{2}$ (for simplicity, we drop in the following the short-distance
cutoff $\alpha$),
\begin{eqnarray}
  S_{\mathrm mix}^{\Phi}=
    -i\int dz_{1}\left(v_{-}^{c}
    \partial_{x_{1}}^{2}\Phi_{-}+i\frac{v_{-}^{p}}{K_{+}u_{+}}\partial_{\tau_{1}}
    \Pi_{-}\right)
  \nonumber\\
    \times\frac{\beta K_{+}}{4\pi}\ln\left[\frac{(x-x_{1})^{2}+u_{+}^{2}\tau_{1}^{2}}{x_{1}^{2}
    +u_{+}^{2}\tau_{1}^{2}}\right]
  \nonumber\\
    -\int dz_{1}dz_{2}\left(v_{-}^{c}
    \partial_{x_{1}}^{2}\Phi_{-}(1)+\frac{i\;v_{-}^{p}}{K_{+}u_{+}}\partial_{\tau_{1}}
    \Pi_{-}(1)\right)
  \nonumber\\  
     \times\frac{K_{+}}{8\pi}\ln\left[x_{12}^{2}+u_{+}^{2}\tau_{12}^{2}\right]
     (1\leftrightarrow 2),
   \label{eq:SmixPhi}
\end{eqnarray}
where $(1)=(x_{1},\tau_{1})$, $x_{12}=x_{1}-x_{2}$,
$\tau_{12}=\tau_{1}-\tau_{2}$, and $dz_{1}=dx_{1}d\tau_{1}$.
The average is taken with the SG action resulting from (\ref{eq:HSG}).
Since the $\Phi_{-}$ field is unpinned, integration over $\Phi_{-}$ gives
corrections $\propto (v_{-}^{c})^{2}$ (and higher order) to the exponent of $x$.

Next, we carry out a canonical transformation from $(\Phi_{-},\Pi_{-})$ to
$(\hat{\Phi}_{-},\hat{\Pi}_{-})=(\theta_{-},\partial_{x}\Phi_{-})$
(the transformation
is canonical, since it preserves the commutation relations and the integration measure
is not changed, because the Jacobian of the transformation has determinant one).
Rewriting $S_{\mathrm mix}^{\Phi}$ in terms of the new fields then allows us to carry
out the integration over $\hat{\Pi}_{-}$. The $\hat{\Pi}_{-}^{2}$ part has the form,
\begin{eqnarray}
  -\frac{u_{-}}{2K_{-}}\int dz_{1}\hat{\Pi}_{-}(1)\left\{\hat{\Pi}_{-}(1)
    +\frac{K_{+}K_{-}}{4\pi u_{-}}\left(v_{-}^{c}\right)^{2}\right.
   \nonumber\\
   \left.\times\int dz_{2}\hat{\Pi}_{-}(2)\partial_{x_{1}}\partial_{x_{2}}
     \ln\left[x_{12}^{2}+u_{+}^{2}\tau_{12}^{2}\right]\right\}.
   \label{eq:PiSquared}
\end{eqnarray}
The inverse of the above operator on $\hat{\Pi}_{-}$ can be expanded
in a power series in $(v_{-}^{c})^{2}$. The linear part in $\hat{\Pi}_{-}$ reads
\begin{eqnarray}
  i\int dz_{1}\hat{\Pi}_{-}(1)\left\{\frac{\beta K_{+}v_{-}^{c}}{4\pi}
    \partial_{x_{1}}\ln\left[\frac{(x-x_{1})^{2}+u_{+}^{2}\tau_{1}^{2}}{x_{1}^{2}
    +u_{+}^{2}\tau_{1}^{2}}\right]\right.
  \nonumber\\
  +\partial_{\tau_{1}}\hat{\Phi}_{-}(1)
  \nonumber\\
  \left.
  +\frac{v_{-}^{p}v_{-}^{c}}{4\pi u_{+}}
    \int dz_{2}\partial_{x_{1}}\ln\left[x_{12}^{2}
      +u_{+}^{2}\tau_{12}^{2}\right]\partial_{x_{2}}
      \partial_{\tau_{2}}\hat{\Phi}_{-}(2)\right\}.
\end{eqnarray}
Carrying out the integration over $\hat{\Pi}_{-}$
we obtain the following contribution to the correlation function in $(v_{-}^{c})^{2}$,
\begin{equation}
  -\frac{K_{-}}{2u_{-}}\left(\frac{\beta K_{+}v_{-}^{c}}{4\pi}\right)^{2}
    \int dz_{1}\left\{\partial_{x_{1}}\ln\left[\frac{(x-x_{1})^{2}+u_{+}^{2}\tau_{1}^{2}}
    {x_{1}^{2}+u_{+}^{2}\tau_{1}^{2}}\right]\right\}^{2}.
\end{equation}
The integral is equal $8\pi(\ln x)/u_{+}$. Including all orders in $v_{-}^{c}$,
the correlation function finally decays $\propto x^{-\gamma_{c}}$, where
\begin{equation}
  \gamma_{c}=\frac{\beta^{2}K_{+}}{2\pi}
    \frac{1}{1-\frac{K_{+}K_{-}}{2u_{+}u_{-}}\left(v_{-}^{c}\right)^{2}}.
  \label{eq:ChargeExp}
\end{equation}

Similarly, the pairing correlation function takes the form
\begin{equation}
  \left\langle e^{i\beta\theta_{+}(x)}e^{-i\beta\theta_{+}(0)}\right\rangle
    \propto x^{-2\beta^{2}/(4\pi K_{+})}
    \left\langle e^{S_{\mathrm mix}^{\theta}(x)}\right\rangle_{\mathrm SG},
\end{equation}
where again $S_{\mathrm mix}^{\theta}$ depends (nonlocally) on the fields
$\Phi_{-}$, $\Pi_{-}$, also vanishing for $v_{1}=v_{2}$,
\begin{eqnarray}
  S_{\mathrm mix}^{\theta}=
    \frac{\beta}{2\pi}\int dz_{1}\left(v_{-}^{c}
    \partial_{x_{1}}^{2}\Phi_{-}+i\frac{v_{-}^{p}}{K_{+}u_{+}}\partial_{\tau_{1}}
    \Pi_{-}\right)
  \nonumber\\
    \times\left[\arctan\left(\frac{x_{1}-x}{u_{+}\tau_{1}}\right)
      -\arctan\left(\frac{x_{1}}{u_{+}\tau_{1}}\right)\right]
  \nonumber\\
  -\int dz_{1}dz_{2}\left(v_{-}^{c}
    \partial_{x_{1}}^{2}\Phi_{-}(1)+\frac{i\;v_{-}^{p}}{K_{+}u_{+}}\partial_{\tau_{1}}
    \Pi_{-}(1)\right)
  \nonumber\\  
    \times\frac{K_{+}}{8\pi}\ln\left[x_{12}^{2}+u_{+}^{2}\tau_{12}^{2}\right]
     (1\leftrightarrow 2).
  \label{eq:SmixTheta}
\end{eqnarray}
The logarithm in (\ref{eq:SmixPhi}) and the arcustangent in (\ref{eq:SmixTheta}) 
give after partial integration a similar leading order contribution. The difference
comes from the $iK_{+}$ present in (\ref{eq:SmixPhi}) but not in
(\ref{eq:SmixTheta}). Here, there are only terms $\propto (v_{-}^{c})^{2}$.
The combination $\partial_{x}\arctan()\partial_{x}\ln()$ does not give
logarithmic contributions after integration.
A similar calculation as above leads to a decay
$\propto x^{-\gamma_{p}}$, where
\begin{equation}
  \gamma_{p}=\frac{\beta^{2}}{2\pi K_{+}}
    \left[1-\frac{K_{+}K_{-}}{2u_{+}u_{-}}\left(v_{-}^{c}\right)^{2}\right].
  \label{eq:PairExp}
\end{equation}

In both cases, the remaining part in the $\hat{\Phi}_{-}$ fields is
either real (and does therefore pin the field) or it is imaginary
but multiplied with derivatives of logarithmes beeing strongly
peaked at $\tau_{1}=0$ and $x_{1}=x$ or $x_{1}=0$,
resulting in a effective contribution
$\propto i(\hat{\Phi}_{-}(x)-\hat{\Phi}_{-}(0))$. In both
cases, the remaining part then becomes a constant ($\neq 0$) for large $x$.

Comparing (\ref{eq:ChargeExp}) with (\ref{eq:PairExp}) we see
that one can express both correlation function in terms of a
single exponent,
\begin{equation}
  \gamma=\frac{K_{+}}{1-\frac{K_{+}K_{-}}{2 u_{+}u_{-}}
    \left(v_{-}^{c}\right)^{2}},
\end{equation}
such that the charge density correlation function decays
$\propto x^{-\beta^{2}\gamma/2\pi}$ and the 
pairing correlation function $\propto x^{-\beta^{2}/(2\pi\gamma)}$.
In our case, $\beta^{2}=2\pi$.

\end{multicols}


\begin{thebibliography}{99}
%
\bibitem{bib:Luttinger}J.M. Luttinger, J. Math. Phys. {\bf 4}, 1154 (1963).
\bibitem{bib:Haldane}F.D.M. Haldane, J. Phys. C{\bf 14}, 2585 (1981).
\bibitem{bib:Jerome}D. J\'{e}rome and H.J. Schulz, Adv. Phys. {\bf 31}, 299 (1982).
\bibitem{bib:Anderson}P.W. Anderson, Science {\bf 235}, 1196 (1987).
\bibitem{bib:DRS}E. Dagotto, J. Riera, and D.J. Scalapino, Phys. Rev. B{\bf 45},
5744 (1992).
\bibitem{bib:RGS}T.M. Rice, S. Gopalan, and M. Sigrist, Europhys. Lett.
{\bf 23}, 445 (1993).
\bibitem{bib:DagRice}For a review, see E. Dagotto and T.M. Rice, Science {\bf 271}, 618 (1996)
and E. Dagotto, cond-mat/9908250.
\bibitem{bib:NWS}R.M. Noack, S.R. White, and D.J. Scalapino,
Phys. Rev. Lett.{\bf 73}, 882 (1994).
\bibitem{bib:Hal}C.A. Hayward \emph{et al.}, Phys. Rev. Lett. {\bf 75}, 926 (1995).
\bibitem{bib:KKA}K. Kuroki, T. Kimura, and H. Aoki, Phys. Rev. B{\bf 54},
R15641 (1996).
\bibitem{bib:UNA}M. Uehara \emph{et al.}, J. Phys. Soc. Jpn {\bf 65},
2764 (1996).
\bibitem{bib:LW}E.H. Lieb and F.Y. Wu, Phys. Rev. Lett. {\bf 20}, 1445 (1968).
\bibitem{bib:BB}P.A. Bares and G. Blatter, Phys. Rev. Lett. {\bf 64}, 2567
(1990).
\bibitem{bib:Fabrizio}M. Fabrizio, Phys. Rev. B{\bf 48}, 15838 (1993).
\bibitem{bib:KR}D.V. Khveshchenko and T.M. Rice, Phys. Rev. B{\bf 50},
252 (1994).
\bibitem{bib:Schu}H.J. Schulz, Phys. Rev. B{\bf 53}, R2959 (1996).
\bibitem{bib:BF}L. Balents and M. Fisher, Phys. Rev. B{\bf 53}, 12133 (1996).
\bibitem{bib:LBF}H. Lin, L. Balents, and M. Fisher, Phys. Rev. B{\bf 56},
6569 (1997).
\bibitem{bib:LBFso8}H. Lin, L. Balents, and M. Fisher, Phys. Rev. B{\bf 58},
1794 (1998).
\bibitem{bib:KLLS}R. Konik \emph{et al.},
cond-mat/9806334 and cond-mat/ 9810332.
\bibitem{bib:Schulzxxz}H.J. Schulz, Phys. Rev. B{\bf 59}, R2472 (1999).
\bibitem{bib:Nerse}A.A. Nersesyan, A. Luther, and F.V. Kusmartsev, Phys. Lett.
A{\bf 176}, 363 (1993).
\bibitem{bib:SLFInfDim}The metal-insulator transition in a spinless two-band
model in infinite dimensions has been studied in Q. Si \emph{et al.}, Phys. Rev.
Lett. {\bf 72}, 2761 (1994) and L. Craco, Phys. Rev. B{\bf 59}, 14837 (1999).
\bibitem{bib:MuEm}K.A. Muttalib and V.J. Emery, Phys. Rev. Lett.
{\bf 57}, 1370 (1986).
\bibitem{bib:Gogolin}A.O. Gogolin, A.A. Nersesyan, and A.M. Tsvelik,
in \emph{Bosonization and Strongly Correlated Systems}
(Cambridge University Press, 1998).
\bibitem{bib:Orignac}The effect of disorder on the different phases has been studied in
E. Orignac and T. Giamarchi, Phys. Rev. B{\bf 53}, R10453 (1996) and
Phys. Rev. B{\bf 56}, 7167 (1997).
\bibitem{bib:Shankar}R. Shankar, Rev. Mod. Phys. {\bf 66}, 129 (1994).
\bibitem{bib:SchulzRev}H.J. Schulz, in \emph{Proceedings of Les Houches Summer School
LXI}, ed. by E. Akkermans \emph{et al.} (Elsevier, Amsterdam, 1995).
\bibitem{bib:LSM}E.H. Lieb, T. Schultz, and D.C. Mattis, Ann. Phys. (N.Y.) {\bf 16},
407 (1961).
\bibitem{bib:YOA}M. Yamanaka, M.Oshikawa, and I. Affleck, Phys. Rev. Lett. {\bf 79},
1110 (1997).
\bibitem{bib:GHR}P. Gagliardini, S. Haas, and T.M. Rice, Phys. Rev. B{\bf 58},
9603 (1998).
\bibitem{bib:Error}The difference of $O_{\mathrm SC}$ to the bosonized form of $O_{S2}$ 
in Ref. \cite{bib:Nerse} is due to an error in \cite{bib:Nerse} (already noted in \cite{bib:Orignac}).
The difference to the fermionic form of $O_{\mathrm SCd}$ in Ref. \cite{bib:Orignac} is due to a mistake in
\cite{bib:Orignac} (private communication).
\bibitem{bib:GExpl}When choosing $g_{x}\gg t$, we obtain $\gamma\rightarrow\infty$ and
$O_{\mathrm SC}\sim$ const (at $T=0$).
\bibitem{bib:NBSZ}A similar finding has been made in numeric calculations (for spinful fermions) in
R.M. Noack \emph{et al.}, Phys. Rev. B{\bf 56}, 7162 (1997).
%
\end{thebibliography}
\end{document}